\documentclass[tightenlines,preprint,aip,jcp,amsmath,amssymb,superscriptaddress,groupedaddress]{revtex4-1}
\usepackage{graphicx}
\usepackage{float}
\usepackage{xr}
\usepackage{mhchem}
\usepackage{makecell}
\usepackage{subcaption}

\usepackage{amssymb,amsmath,amsthm,mathrsfs,comment,afterpage,accents}
\usepackage{fixltx2e}

\makeatletter
\def\@dotsep{4.5}
\makeatother

\usepackage{dcolumn}% Align table columns on decimal point
\usepackage{bm}% bold math

\newcommand{\bra}[1]{\langle #1|}
\newcommand{\ket}[1]{|#1\rangle}
\newcommand{\braket}[2]{\langle #1|#2\rangle}

\newcommand\MLmodelName{bpopNN}

\begin{document}

\title{Incorporating Electronic Information into Machine Learning Potential Energy Surfaces via Approaching the Ground-State Electronic Energy as a Function of Atom-Based Electronic Populations}

\author{Xiaowei Xie}
\affiliation{Department of Chemistry, University of California, Berkeley, California 94720, United States}
\affiliation{Energy Technologies Area, Lawrence Berkeley National Laboratory, Berkeley, California 94720, United States}
\author{Kristin A. Persson}
\affiliation{Department of Materials Science and Engineering, University of California, Berkeley, California 94720, United States}
\affiliation{Energy Technologies Area, Lawrence Berkeley National Laboratory, Berkeley, California 94720, United States}
\author{David W. Small}
\affiliation{Department of Chemistry, University of California, Berkeley, California 94720, United States}
\affiliation{Molecular Graphics and Computation Facility, College of Chemistry, University of California, Berkeley 94720, California United States}
\email{dsmallchem@gmail.com}

\date{\today}

\begin{abstract}
Machine Learning (ML) approximations to Density Functional Theory (DFT) potential energy surfaces (PESs) are showing great promise for reducing the computational cost of accurate molecular simulations, but at present they are not applicable to varying electronic states, and in particular, they are
not well suited for molecular systems in which the local electronic structure is sensitive to the medium to long-range electronic environment.
With this issue as the focal point, we present a new Machine Learning approach called ``\MLmodelName'' for obtaining efficient approximations to DFT PESs.  The methodology is based on approaching the true DFT energy as a function of electron populations on atoms, which may be realized in practice with constrained DFT (CDFT).  The new approach creates approximations to this function with deep neural networks.  These approximations thereby incorporate electronic information naturally into a ML approach, and optimizing the model energy with respect to populations allows the electronic terms to self-consistently adapt to the environment, as in DFT.
We confirm the effectiveness of this approach with a variety of calculations on Li$_n$H$_n$ clusters.
\end{abstract}

\maketitle

\section{Introduction}
DFT computations continue to be the predominant approach to first-principles modeling of chemical systems,\cite{ref61, ref62, ref63} while empirical force-field based approximations continue to dominate large scale simulations.\cite{hansson2002molecular, karplus2002molecular}  As one category of approaches intent on bridging this divide, ML-based methods for approximating first-principles PESs are drawing much attention.  These methods largely reduce the computational cost while remaining remarkably near to the accuracy of quantum chemistry calculations, and practical applications in chemistry, physics and material science have been demonstrated\cite{behler2017first,schmidt2019recent,deringer2019machine,ref65}. So far, predictions of the electronic energy for neutral, closed-shell molecules (equilibrium or off-equilibrium) have been successfully demonstrated.  A ubiquitous theme in such methodologies is the development of translationally, rotationally, and permutationally invariant descriptors based on nuclear positions. Prominent examples include the atom-centered symmetry functions\cite{behler2007generalized,behler2011atom} of Behler and Parrinello and the smooth overlap of atomic positions (SOAP)\cite{bartok2013representing} by Bartok et al. Some specific ML models that fit the general positions-to-energy characterization are RuNNer,\cite{behler2018runner} Tensormol,\cite{yao2018tensormol,yao2017many,herr2018metadynamics,yao2017intrinsic,herr2019compressing} SchNet,\cite{schutt2018schnet, schutt2019unifying} ANI-1,\cite{smith2017ani, smith2019approaching, smith2018less} AMP,\cite{khorshidi2016amp} DeepMD,\cite{wang2018deepmd} LASP,\cite{huang2019lasp} QML,\cite{ref66, ref78} PhysNet, \cite{unke2019physnet} GAP,\cite{bartok2015g} models based on atom-density representations by Ceriotti and coworkers,\cite{grisafi2018transferable,grisafi2018symmetry,grisafi2019incorporating,de2016comparing,willatt2018feature} the spectral neighbor analysis potential (SNAP), \cite{thompson2015spectral} graph kernel methods,\cite{tang2019prediction,ferre2017learning} graph neural networks, \cite{ref57,chen2019graph} methods based upon eigenvalues of the Coulomb matrix, \cite{rupp2012fast, hansen2013assessment} Bag-of-Bonds, \cite{hansen2015machine} permutation invariant polynomials,\cite{shao2016communication} and others. Impressive examples of large-scale applications of these methods include simulating proteins, \cite{yao2018tensormol} amorphous carbon, \cite{ref67,sosso2018understanding,caro2018growth} constructing phase diagrams of amorphous Li$_x$Si, \cite{ref64} etc.

However, descriptors based on atom positions alone exhibit some important limitations. For example, they cannot be applied to more than one overall charge or spin state. Furthermore, far-away changes in the chemical environment may induce changes in the local charge or spin state of a region of a molecule even if the local geometry is not changed appreciably.  We illustrate this with an example in section \ref{long-range} below.
This general problem applies to many molecules and materials.  Examples relating to conjugation, like the emergence of polyradicality with increasing length in acenes,\cite{BendikovJACS2004,HachmannJCP2007,JiangJPCA2008} armchair versus zigzag edge effects in graphene nanoribbons,\cite{DuttaJMC2010,CervantesPRB2008,BaroneNL2006} and the gradual ascencion to the effective conjugation length in a variety of oligomers,\cite{MeierAP1997,IzumiJACS2003,KishinoPRB1998} come to mind.  Even more marked are examples involving explicit ionization and charge transfer (CT).  These include outer-sphere electron transfer processes such as those occuring in Ferredoxin protein cores and other proteins\cite{Gray1994,MarcusBBA1985} 
and large-scale organic donor-acceptor complexes with appealing electronic properties.\cite{JeromeCR2004,ForrRPP2001,TakahashiJPCC2012}

We stress the need for advanced ML-based methods to describe complicated heterogeneous systems such as found in electrode surfaces like the solid electrolyte interface (SEI) in lithium-ion batteries (LIB).\cite{peled1979electrochemical,nie2013lithium,winter2009solid,wang2018review,peled2017sei}  As a small glance into the complexities involved here, consider that the ground state of neutral lithium ethylene carbonate, a key intermediate in the reductive decomposition process to form the LIB SEI, varies in character between neutral and CT, i.e.\ unpaired electron on Li or on the rest of the molecule.  This variation is subtly dependent on where the Li atom locates around the EC molecule;\cite{wang2001theoretical} the transition between the two characters can be fairly rapid, and so for position-based descriptors, whose terms don't change rapidly in such transitions, this situation is clearly a practical difficulty.
This is just one part of the SEI's multiplex of charged and uncharged species and the reactions between them, underscoring significant challenges for ML descriptors.

In principle, these general effects can be captured by a position based descriptor along with the total charge and spin multiplicity of the molecule, as after all these are the only inputs needed for a ground-state calculation.  However, size extensibility, the crucial property that a ML model may be uniformly and consistently applied to systems of different sizes, favors the use of local, typically atom-centered, descriptors, which are difficult to reconcile with global parameters like total charge and spin.  And, stretching these to include the (possibly very) long range information required for these cases is challenging in practice in terms of computational efficiency and numerical stability.

The above issues stem from a lack of explicit electronic information in the descriptors, which suggests a general remedial approach. To successfully incorporate electronic information,  the basic scheme of the parent DFT model should be followed such that electronic terms can adapt according to changes in neighboring electronic information, preferably in a self-consistent or equilibrated way.  

In this context, we would like to briefly reassess some of the models mentioned above.
The SOAP based descriptors make use of smeared atom density distributions, and together these bear a rough resemblance to electron density.  A recent variant\cite{grisafi2019incorporating} uses the electrostatic potential (ESP) associated with such distributions as the basis for the SOAP expansion.  This approach provides enhanced long-range information, and we will comment more on this important development at the end of this paper.  Nevertheless, these density distributions are fixed and do not adapt to the electronic environment. 

Several of the above ML approaches include a separate long-range, pairwise electrostatic contribution to the energy.  This inclusion has been shown to be very useful for describing large Zn$_{n}$O$_{n}$ clusters, \cite{ref75}, water clusters, \cite{yao2018tensormol,wang2018force,morawietz2013full,morawietz2013density} proteins,\cite{yao2018tensormol} and other examples \cite{deng2019electrostatic,shen2016multiscale}.  Popelier and coworkers first proposed to employ NNs to construct environment dependent multipoles.\cite{houlding2007polarizable}  Artrith et.\ al.,\cite{ref75} and later on Yao et.\ al.\ \cite{yao2018tensormol} adopted a scheme where they pretrain a neural network to predict atomic point charges, using the same descriptors (i.e. symmetry functions) as they used to train energies. These charges were subsequently used to compute the electrostatic term through Ewald summation \cite{ewald1921berechnung} for periodic systems, or Coulomb's law for molecules. Another approach fits molecular dipoles to infer partial charges. \cite{gastegger2017machine} We note that there are also independent ML models specialized for predicting partial charges, \cite{bleiziffer2018machine,wang2019graph} multipole moments,\cite{bereau2015transferable, bereau2018non, grisafi2018symmetry}  and even the full electron density. \cite{grisafi2018transferable,chandrasekaran2019solving}. The latter method has been used e.g. to guide the inverse design of chemical materials based on electron density information, to serve as an initial guess for SCF convergence, and more.  The above-mentioned methods all predict the electronic terms directly from local position-based descriptors, and thus are not properly adaptable to the electronic environment. 

In another direction, the various charge equilibration schemes, such as EEM\cite{MortierJACS1985} and QEq\cite{RappeJPC1991} and several newer examples,\cite{NistorJCP2006,ChenCPL2007,ZhangJCPA2009,WilmerJPCL2012,VerstraelenJCP2013,WellsJPCC2015,NaserifarJCP2017} compute atomic charges self-consistently.
These models therefore exhibit some basic flexibility in
the context of the above issues.  However, they also employ very simple forms; this is advantageous for application to very large molecular systems, but it limits the accuracy as compared to the above ML-based models.
Ghasemi, et al\cite{ghasemi2015interatomic} developed a charge equilibration scheme where the intra-atomic energy function comes from a deep NN that is parametrized by DFT data, and they used this to predict the energies for both neutral and ionized NaCl clusters.  To our knowledge, this important development is the only extant positions-to-energy model that employs a high-level ML approximation and that is also able to accommodate varying electronic structure.
However, the model is exclusively trained on self-consistent DFT energies and its charges do not enter the descriptor. Hence the charges are effectively auxiliary parameters as opposed to being more directly associated with the electron density.  This precludes, for example, obtaining more than one SCF solution at a given geometry.

An open question remains as to how much electronic information should be incorporated.  At one extreme, it is possible to utilize the full electron density and create ML models that map this directly to the energy. Some of these efforts are concerned with finding new DFT approaches with unprecedented accuracy.\cite{dick2019machine, ref68, ref73, ref77}  Others pertain more to the fact that commonly used functionals do not provide a direct non-iterative link between density and energy, so a ML approximation for this, as with other orbital-free approaches, can greatly reduce the associated burden.\cite{ref69, ref70, ref72, ref71, ref77, brockherde2017bypassing, schutt2019unifying}  Effective self-consistency is then obtained by optimizing the energy with respect to the density.  This level of electronic information is ideal in terms of accuracy, and it easily addresses the above issues, but it entails a significant increase in computational requirements compared to the above models.

In the present paper, we attempt to identify the simplest amount of electronic information that is sufficient to solve the above issues, and how to incorporate it effectively.  In a sense, we are seeking a good balance between the above position-only approaches and the full-density based approach.  We focus on approximating regular DFT functionals, and we explore the incorporation of atomic electron populations and associated electrostatic interaction terms into the descriptor.  Effectively, this divides the global charge and spin parameters into local terms that may be naturally incorporated into local descriptors.  In the following sections, we exploit CDFT to realize a natural map from intrinsic populations (self-consistently optimal ones or otherwise) to energies, which may be approximated with deep neural networks, leading to a simple self-consistent approach that is qualitatively true to the parent DFT functional.

\section{Theory}
\subsection{Target Energy Function}

\subsubsection{Formal viewpoint}

In the Levy constrained search approach in DFT,\cite{LevyPNAS1979} each density $\rho$ maps to the lowest possible energy $E[\rho]$ obtainable from a many-electron wave function 
associated with that density.  Thinking of the density as being a reduction of the many-electron density matrix, the constrained-search idea may be 
generalized to any collection of reduced variables $\{v_i\}$ (which may be functions, etc.):

\begin{equation}\label{basicLevyCS}
E[\{v_i\}] = \min_{\Psi \rightarrow \{v_i\} } \bra{\Psi} \mathbf{H} \ket{\Psi},
\end{equation}
where $\mathbf{H}$ is the many-electron Hamiltonian.  Of course, in practice, the search over wave functions and the energy expectation value on the right 
hand side will be replaced by a search over orbitals and an approximate functional, respectively.

For $\{v_i\}$ in this paper, we are primarily concerned with reducing the density to atom populations, in
particular Becke populations.\cite{BeckeJCP1988}  For these, the density is multiplied by a weight function and integrated to produce a population value.  For atom $i$ and spin $\sigma$ we have

\begin{equation}\label{basicBeckePop}
p_{i,\sigma} = \int w_i(\mathbf{r}) \rho_{\sigma}(\mathbf{r}).
\end{equation}
The weight functions $w_i$ are localized to their respective atoms, they take on values between 0 and 1, maximizing on the pertinent atom, and they sum over 
all atoms to 1.

Our target function for machine learning is essentially $E[\bm{p}]$, where we have placed the populations into a vector $\bm{p}$.
Technically, this function is defined only for population values that are non-negative and sum to appropriate values for the numbers of 
electrons of each spin, or equivalently, the total charge and spin
multiplicity.  We can extend this function to the larger domain of non-negative population vectors with no sum conditions by first applying 
$P_{N_{\alpha},N_{\beta}}$, which ``projects'' the population vector to a vector obeying the sum conditions:

\begin{equation}\label{defPqm}
[P_{N_{\alpha},N_{\beta}}(\bm{p})]_{i,\sigma} = p_{i,\sigma} + \frac{N_{\sigma} - \sum_j p_{j,\sigma}}{N_{\sigma}},
\end{equation}
where $N_{\alpha}$ and $N_{\beta}$ are the numbers of $\alpha$ and $\beta$ electrons, respectively.  Hence, our formal target function is
$E[P_{N_{\alpha},N_{\beta}}(\bm{p})]$.

The target function may be optimized with respect to the population values, with
the latter considered as input variables.  This is much the same as optimizing DFT energies with respect to the density or orbitals.  Optimizing with respect to
the populations thus defines an SCF procedure.  As to the issue of representability here, we only need to ensure that the populations (after projection) are 
non-negative.  

At this point, we need a way to obtain practical approximations to this target function, which is the subject of the next subsection.

\subsubsection{The target function in practice: CDFT}

The above ideas may be applied to any of the functionals used in common practice, e.g. B3LYP as we use below. That is, they each may be formally reduced to a functional of populations.  

Training data for the target function must be obtained with non-standard DFT calculations. Fully self-consistent DFT solutions provide data for certain population choices, however, by itself this would constitute a very limited training set.
Energies for other population choices may be obtained by using constrained DFT (CDFT),\cite{DederichsPRL1984,WuPRA2005,KadukCR2012} which uses Lagrange multipliers to optimize the DFT energy
under various constraints.  The use of constraints on Becke-weight based populations has been implemented in Q-Chem,\cite{shao2015advances} and allows us to, in principle, obtain
the energy for any given set of populations.  Much as how there are generally numerous wave functions (and thereby energies)
associated with a given density, there will overall be many energies associated with a specific population vector.  In fact, the energy spread is much wider for 
populations
than for the density, because there will generally be many densities associated with a given population vector.  As with standard DFT calculations, care must be 
exercised to attempt to obtain the lowest energy CDFT solution.

The accuracy of ML models of the DFT energy is significantly enhanced by incorporating energy derivatives into the loss function, i.e.\ it is desirable
to train the model towards both the direct value of the target function and its first derivatives.  The $xyz$ position derivatives for CDFT have been 
previously derived.\cite{WuJPCA2006}  For this present paper, we also need a way to compute the derivatives of the 
CDFT energy with respect to populations.  This is described in the Appendix.

\subsection{Model Energy Function}

\begin{figure}[H]
  \centering
  \includegraphics[width=\linewidth]{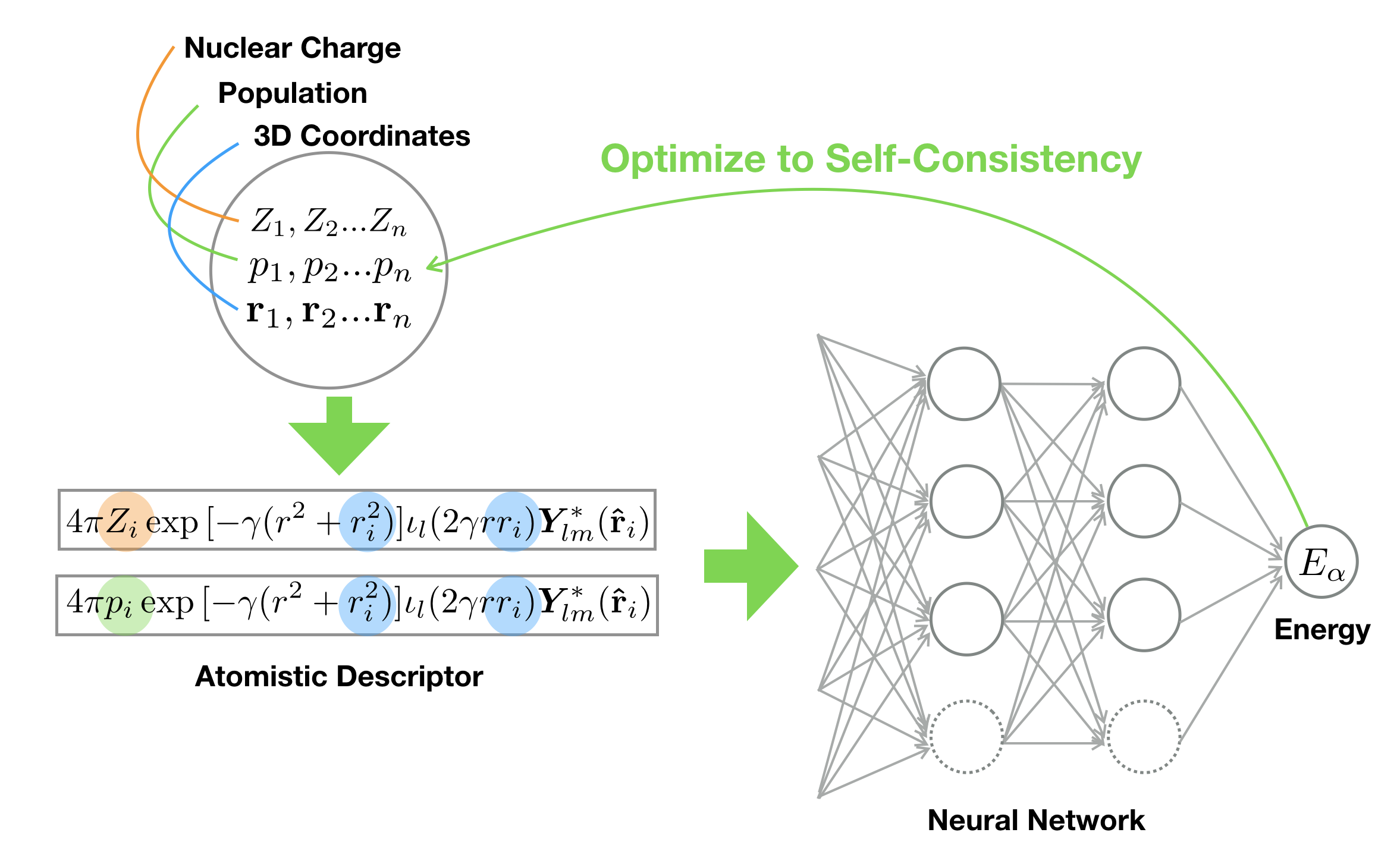}
  \caption{General concept for the ML potential}
  \label{fig:concept}
\end{figure}

We adopted a widely accepted scheme in the ML PES field, where the total energy is decomposed into atom-centered contributions to ensure transferability across systems with different sizes and compositions. A deep neural network (NN) framework was used to map an atomistic descriptor 
$D_i[\bm{Z}, \bm{p}, \bm{r}]$ to the atomic energy. The model energy function can be written as 
\begin{equation}
E_{\text{ML}} = \sum_{\mu} \sum_{i\in A_{\mu}}\epsilon_{\mu}[D_i[\bm{Z}, \bm{p}, \bm{r}]] + E_{\text{intra}}[\bm{Z}, \bm{p}, \bm{r}] + E_{\text{el}}[\bm{Z}, \bm{p}, \bm{r}],
\end{equation}
where $\epsilon_{\mu}$ represents the output from the atomic neural network for the element $\mu$, and this function is applied to the descriptor for atom i;
$A_{\mu}$ is the set of atom indices corresponding to element $\mu$; $\bm{Z}$ and $\bm{r}$ contain the atomic numbers and coordinates, respectively, for the entire molecule.  $E_{\text{intra}}$ and $E_{\text{el}}$ are simple intra-atomic and pairwise electrostatic terms, respectively, and will be described below. An overview of our model, which we will call ``\MLmodelName'' for ``Becke Population Neural Network'', is shown in Figure ~\ref{fig:concept}. 
More details are given in the following subsections.

\subsubsection{Atomistic Descriptor}

The descriptor we use in this paper is essentially a variant of SOAP,\cite{bartok2013representing} modified to describe the electronic as well as nuclear environments. 
The regular SOAP descriptor is based on an atomic power spectrum of a basic density distribution that represents the nuclei: for each atom, the latter is expanded
in a set of radial basis functions (RBF) and spherical harmonics that are symmetric about that atom, and the coefficients of this expansion directly produce the spectrum. In our case, we want to adapt the density distribution to represent the varying electrostatic environment.

The basic idea is to use the ESPs of the nuclei, and the ESPs of the $\alpha$ populations treated as point charges, and likewise for the $\beta$ populations.  In fact, as in SOAP, we use separate distributions for each element type, since this makes for a more refined descriptor.  In our lithium hydride applications below, we simplified the descriptor to be based only on the nuclear charges and the total electron population on each atom.  In the present subsection, we will continue the development using spin populations.  This is more general, and reducing this descriptor to the total-population one is straightforward.

To simplify the computation of the descriptor elements, we approximate the $1/r$ term in the ESPs with a single Gaussian function.  Hence our density distributions take the form
\begin{equation}\label{a}
\rho_{\mu,t}(\mathbf{r}) = \sum_{i \in A_{\mu}} \chi_i^t B \exp{(-\gamma \left| \mathbf{r} - \mathbf{r}_i \right|^2)},
\end{equation}
where $t$ is either $\nu$ (for nuclear), $\alpha$, or $\beta$ corresponding to the type of point population, and thus $\chi_i^{\nu} = Z_i$,
$\chi_i^{\alpha} = p_{i,\alpha}$, and $\chi_i^{\beta} = p_{i,\beta}$.
Here we used $B=1.128$ and $\gamma=0.171$, and $r$ is in the unit Bohr radii.

To build the descriptor for any atom in the molecule, we set the origin to that position and project the radial slices of $\rho_{\mu,t}(\mathbf{r})$ onto 
the pertinent set of spherical harmonics $\bm{Y}_{lm}(\mathbf{\hat{r}})$: 
\begin{equation}
\rho_{\mu,t}(\mathbf{r}) = \sum_{i \in A_{\mu}} \sum_{lm} c_{ilm}^{t}(r) \bm{Y}_{lm}(\mathbf{\hat{r}}).
\end{equation}
Note that the $i$ sum includes the central atom (if it is of the element type $\mu$).
In practice, the $l$ sum must be limited to a range.  In the applications below, we use a maximum $l$ value of 6.
Since $\chi_i^t$ does not depend directly on positions, the analytical form of the radial terms can be derived following previous works \cite{bartok2013representing, kaufmann1989single} as
\begin{equation}
c_{ilm}^{t}(r) \equiv 4 \pi \chi_i^t B \exp{[-\gamma(r^2+r_i^2)]} \iota_{l}(2 \gamma r r_i) \bm{Y}_{lm}^{*}(\mathbf{\hat{r}}_i),
\end{equation}
where $^*$ denotes complex conjugation and $\iota_{l}$ are the modified spherical Bessel functions of the first kind.  Summing over pertinent atoms gives
\begin{equation}
c_{lm}^{\mu,t}(r) = \sum_{i \in A_{\mu}} c_{ilm}^{t}(r),
\end{equation}
and projecting these onto RBFs gives a sequence of expansion coefficients $c_{nlm}^{\mu,t}$ with an extra index $n$.  Underlying the projections is the evaluation of the integrals $\braket{g_{n} \bm{Y}_{lm}}{\rho}$, with $g_n$ denoting an RBF.  In general, this can be done by either analytical or numerical integration (e.g. Gauss-Legendre quadrature), 
but in our case, we bypass these complications by simply evaluating $c_{lm}^{\mu,t}(r)$ on a grid of $r$ values.
The resulting coefficients are smooth with respect to nuclear positions, and technically, one could choose RBFs whose projection coefficients match the evaluated grid values.  Hence the grid spacings can in principle be selected to produce a result that is as satisfactory as the explicit RBF/integration approach. A simple linear grid between 0.5 and 13.2 \AA{} was used in this work; further refinement of the grid will be addressed in the future. 

At this point, one can construct the power spectrum by contracting over $m$, 
\begin{equation}
p_{nl}^{\mu,t} = \sum_{m=-l}^{l} \| c_{nlm}^{\mu,t} \|^2
\end{equation}
It is easy to prove the rotational invariance of $p_{nl}^{\mu,t}$, as 3D rotations correspond to transformations of blocks of the $c_{nlm}^{\mu,t}$ by Wigner matrices, which are unitary.

For each atom in the molecule, we employ a separate power spectrum for each element-type in the molecule, i.e.\ the power spectrum is split into different element channels. There is active research aimed at reducing the cost associated with adding more atomic species,\cite{herr2019compressing, willatt2018feature, artrith2017efficient, ref78} which is also of future interest for us. For now, the above treatment is sufficient to prove the principle.

A radial cutoff of 13.2 \AA{} was used below in computing the descriptor, i.e.\ atom pairs beyond this cutoff were neglected.  The reason for this choosing this relatively long cutoff is that, as described below, in addition to the NN energies, we include a simple long-range pairwise electrostatic term, as has been shown to be very useful in previous papers.\cite{ref75, yao2018tensormol}  The influence of the switch in dominance between short-range NN and explicit electrostatic terms is a general issue, but because this question is separate from the main theme of this work, we chose the long cutoff to minimize this influence.

\subsubsection{Neural Network Architecture}

We employ separate neural networks for different element types, and for the applications below, each element network contains two hidden layers with 30 neurons in each layer.  This size of neural network was used to best accommodate the number of training data and prevent overfitting. Nonlinearities were introduced using a modified softplus activation function $\ln{(1+\exp{(100x)})}/100$.  This form resembles the RELU activation function, which itself is highly popular because it overcomes the vanishing gradient problem,\cite{ref76} yet this softplus variant has the advantage of being twice continuously differentiable.

In previous works, the ML model is usually trained towards the DFT atomization energy, as opposed to total energy. This has been shown to accelerate and balance training.\cite{yao2018tensormol}  The definition of atomization energy is ambiguous in the case of charged molecules, hence we select a quadratic function of atomic charge that best fits the DFT energies for the pertinent charge states of the isolated atom.  $E_{\text{intra}}[\bm{Z}, \bm{p}, \bm{r}]$ is then the sum of these energies for each atom.  For lithium and hydrogen atoms found in the molecular applications below, we used the following quadratic functions as baseline for atomic energies

\begin{equation}
E_{\text{intra}, \text{Li}, i} = 0.0901 q_{i}^{2} + 0.1088 q_{i} - 7.4787
\end{equation}
\begin{equation}
E_{\text{intra}, \text{H}, i} = 0.2471 q_{i}^{2} + 0.2403 q_{i} - 0.4805,
\end{equation}
where $q_{i}$ is the atomic partial charge for atom $i$, i.e.\ $q_{i} = Z_{i} - p_{i,\alpha} - p_{i,\beta}$. This resembles the intra-atomic term in charge equilibration models. \cite{ghasemi2015interatomic}

We also used a simple pairwise coulomb term to incorporate long-range electrostatics, as shown in the following equation.
\begin{equation}
E_{\text{el}}[\bm{Z}, \bm{p}, \bm{r}] = \sum_{j > i} \mathrm{tanh} (\kappa r_{ij}) \frac{ q_{i} q_{j} } { r_{ij} }
\end{equation}
Parameters $\kappa$ (one each for different element pair types) were trained together with other parameters in the neural network.

The loss function for the training is 
\begin{equation}\label{lossfxn}
\mathcal{L} = \sum_{M} \dfrac{(E_{M}^{\text{CDFT}}-E_{M}^{\text{ML}})^2}{N_{\text{atom},M}} + \gamma_{1} \sum_{M} \dfrac{\|\mathbf{F}_{M}^{\text{CDFT}}-\mathbf{F}_{M}^{\text{ML}}\|^2}{N_{\text{atom},M}} + \gamma_{2} \sum_{M} \dfrac{\|\mathbf{f}_{M}^{\text{CDFT}}-\mathbf{f}_{M}^{\text{ML}}\|^2}{N_{\text{atom},M}}.
\end{equation}
For the model presented below, both $\gamma_1$ and $\gamma_2$ were set to 1.5.
The above sums go over the training molecules, and $\mathbf{F}$ is the position gradient, $\mathbf{f}$ is the population gradient, and $N_{\text{atom},M}$ represents the number of atoms in molecule $M$. Adaptive moment solver (Adam)\cite{kingma2014adam} was used to update the NN weights during training.  Our implementation is built up from the open-source package TensorMol,\cite{yao2018tensormol} which takes advantage of the automatic differentiation scheme of Tensorflow.\cite{tensorflow2015-whitepaper}

\section{Results and Discussion}
We here test the performance of our method on Li$_n$H$_n$ clusters of varying size and geometry, and overall charges of +1, 0, and -1. We chose this for the simplicity of having only two element types. Lithium hydride clusters have been studied in several theoretical\cite{RaoJPC1986,Sapse1995,ChenJPCA2005,NolanPRB2009,SatoPRA2015,HuangSR2016}
and experimental\cite{AntoineJCP1996,WangJPCA2007} works, and have implications in various applications including hydrogen storage.\cite{WagnerPCCP2012,HarderCC2012,StaschACIE2014,HuangIJHE2015,WangIJHE2016}

In the calculations below, we employ the descriptor described above, but, as mentioned above, we modified it so that the total electron population on each atom, rather than the 2 spin populations, is used for the electronic terms.

\subsection{Training-data generation and model training}

In the training dataset, the cluster sizes vary from from $n=1$ to 24, with increments of 3 or 4.  We use two basic types of structures from which we obtain more samples: (1) collections of relatively widely separated LiH molecules (2-5  \AA {} between molecules), and (2) denser Li$_n$H$_n$ systems with roughly cubic structures with alternating Li and H.  We denote the first dataset as ``sparse'', and the second as ``non-sparse'' from here on.

For the sparse data, structures from DFT geometry-optimization trajectories were used in the training dataset.  For the non-sparse data, structures both from geometry-optimization trajectories and Ab Initio Molecular Dynamics (AIMD)\cite{marx_hutter_2009} trajectories were used.  The AIMD simulations used a temperature of 500 K and a time step of 20 a.u.\ ($\approx$ 0.5 fs).

All structures were generated using overall charge 0, and calculations for the ionized molecules used the same geometries.
During the geometry optimizations, the LiH units in the sparse structures cluster together in various ways and become less sparse, although they remain less dense than the non-sparse data.  All of these geometries exhibit alternating Li and H atoms, i.e.\ no structures with Li clusters nor \ce{H2} molecules were obtained.  This is because Li clusters often entail spin polarization, and therefore we reserve the generation of such structures for future work in which spin populations are used.  More details of the geometries used in the training dataset are included in the SI.

All DFT computations were conducted using the QChem\cite{shao2015advances} program.  The geometry-optimization trajectories for the sparse data used the B3LYP\cite{BeckeJCP1993,StephensJPC1994} functional with D3(BJ) dispersion\cite{grimme2011effect}, the AIMD trajectories for the non-sparse data used the $\omega$B97X-D\cite{chai2008long} functional, and the geometry-optimization trajectories for the non-sparse data used B3LYP (without dispersion).  The reason for the overall use of different functionals is that we began with $\omega$B97X-D, then switched to using B3LYP-D3(BJ) for computational efficiency, and then encountered some implementational problems with D3(BJ) for Li-Li interactions, so we finally switched to using B3LYP alone.  These variations only occurred for the generation of structures; for all geometries used in the training and test sets (the latter is described below), we (re-)computed the DFT energy using B3LYP.  All DFT calculations in this paper used the def2-SVPD basis set.\cite{RappoportJCP2010}

For the final training set, we selected one per every several geometries from each trajectory, with an interval of 2 to 7 depending on the trajectory's length.  On each of these geometries, we computed unconstrained B3LYP solutions for the considered overall charges (+1,0,-1), and for each of these three,
we performed 10 separate CDFT calculations.  For each of these, we added a random (uniform distribution from -0.05 to 0.05) charge fluctuation to each atom's charge value from the pertinent unconstrained SCF solution, and then computed the associated CDFT energy, nuclear forces, and charge forces (see Appendix for the latter).

The training data ($\sim 37000$ points) were randomly separated to a training (95\%) and a validation set (5\%) at the outset, and the validation error provides an estimate on how well the model is trained.  Typically the model reaches its best performance on test sets after $\sim 500$ epochs of training. The mean absolute error (MAE) on the independent validation set of the energy is 1.523 kcal/mol per molecule.

\subsection{Testing the model}

\subsubsection{Li$_{29}$H$_{29}$ and Li$_{32}$H$_{32}$ clusters}

We generated two test sets with larger molecules to test the transferability of our model: non-sparse \ce{Li32H32} (+1,0,-1 charge) from an AIMD 
trajectory and sparse \ce{Li29H29} (+1,0,-1 charge) from a geometry-optimization trajectory.  For those test sets, CDFT calculations for off-equilibrium charges were not conducted as we focus on comparing the self-consistent solutions from DFT and the \MLmodelName{} model.

As discussed earlier, one advantage of our model is that the populations (i.e.\ partial charges at present) can be optimized self-consistently.  In other words, the ML-model energies are optimized with respect to the partial charges.  This process is reminiscent of the SCF procedure in DFT calculations, and we denote this charge optimization procedure as ``SCF-q'' in the remaining text.  In general, each of these types of SCF can produce more than one stationary point, i.e.\ solution.  Since our objective is to find counterpart solutions between DFT and ML and then compare them, we used the actual DFT partial charges as initial guesses for SCF-q calculations for each geometry in the \ce{Li29H29} and \ce{Li32H32} test sets.  
Other initial guesses can be used but this increases the odds of obtaining an SCF-q solution that does not correspond to the SCF one even when a pair of corresponding solutions exist.  Of course, general and effective initial guess schemes will eventually need to be developed for the partial charges. The Li partial charges in our systems usually exhibit values between 0.5 and 0.8 (although the actual range covered throughout the whole dataset is $-0.31$ to $0.99$).  It turned out that we obtained the same SCF-q solutions when starting with uniform guesses of 0.4 for Li charges and -0.4 for H charges as when starting from DFT partial charges, demonstrating the model's capability of dealing with general reasonable initial guesses for partial charges.  

\begin{figure}[H]
     \centering
     \includegraphics[width=\linewidth]{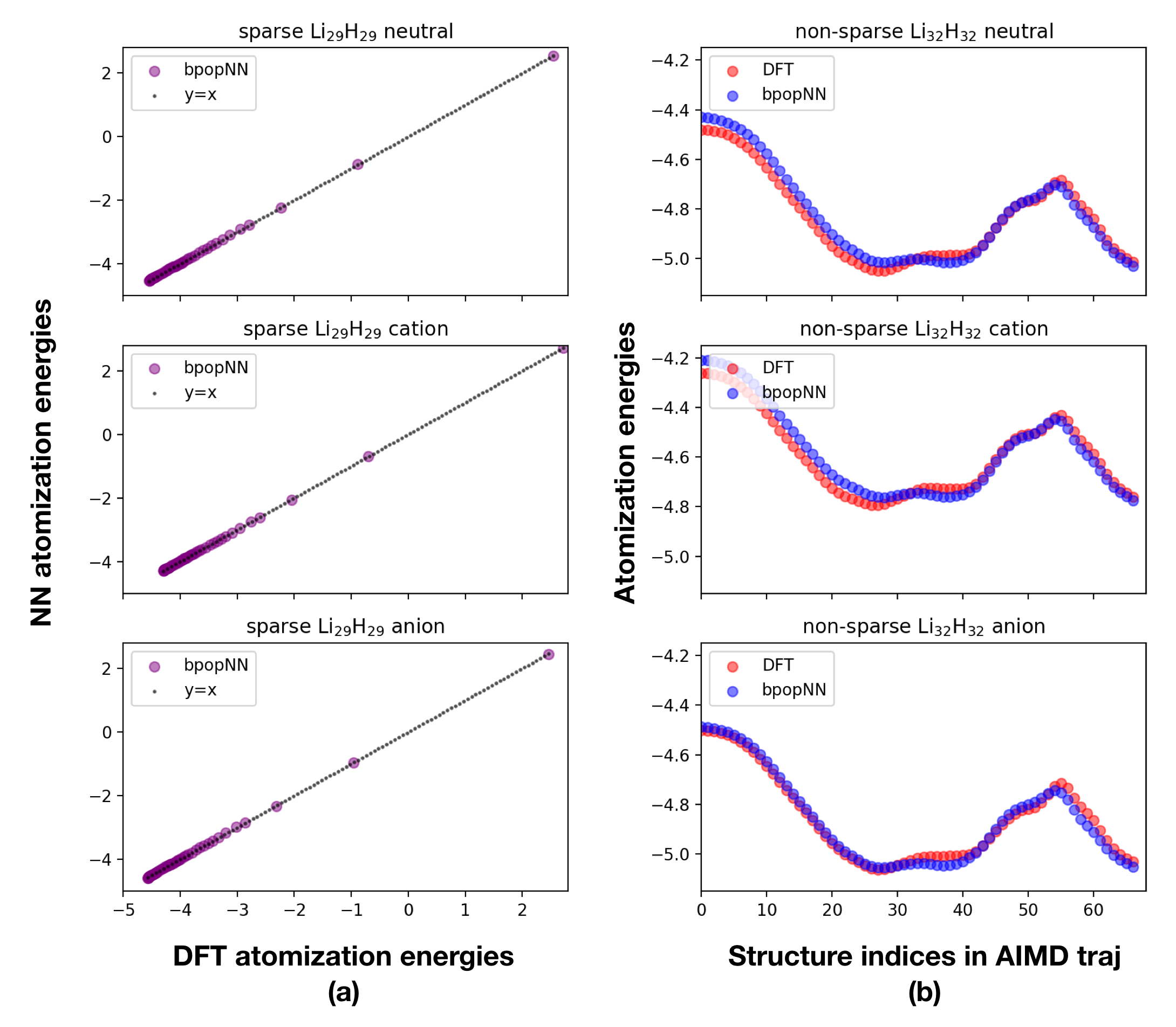}
     \caption{Performance on sparse \ce{Li29H29} and non-sparse \ce{Li32H32} test sets.  All energies are in a.u.}
     \label{fig:fig1}
   \end{figure}

In Figure ~\ref{fig:fig1}, we plot a comparison between the SCF-q \MLmodelName{} energies (converged to threshold $10^{-7}$ Hartree/e) and the DFT energies for neutral and ionized molecules. The statistics for energy and charge errors are shown in Table ~\ref{table:table1}. For the sparse \ce{Li29H29} test set (Figure ~\ref{fig:fig1} (a)), the \MLmodelName{} energies are showing close agreement with the DFT reference energies, with a charge MAE of $\sim 0.01$ per atom, an energy MAE of $\sim 0.15$ kcal/mol per atom. Non-sparse \ce{Li32H32} (Figure ~\ref{fig:fig1} (b)) provides a harder test set - there are more neighbors around each center atom and the Li-H bonds are actively forming and breaking in the AIMD trajectory, which makes it more challenging for the descriptor. We still observe a charge MAE of $\sim 0.01$ and energy MAE of $\sim 0.2$ kcal/mol per atom for those cases. There is no obvious deterioration in performance on ionized molecules compared with neutral molecules. 

\begin{figure}[H]
  \centering
  \begin{subfigure}[b]{0.8\linewidth}
    \includegraphics[width=\linewidth]{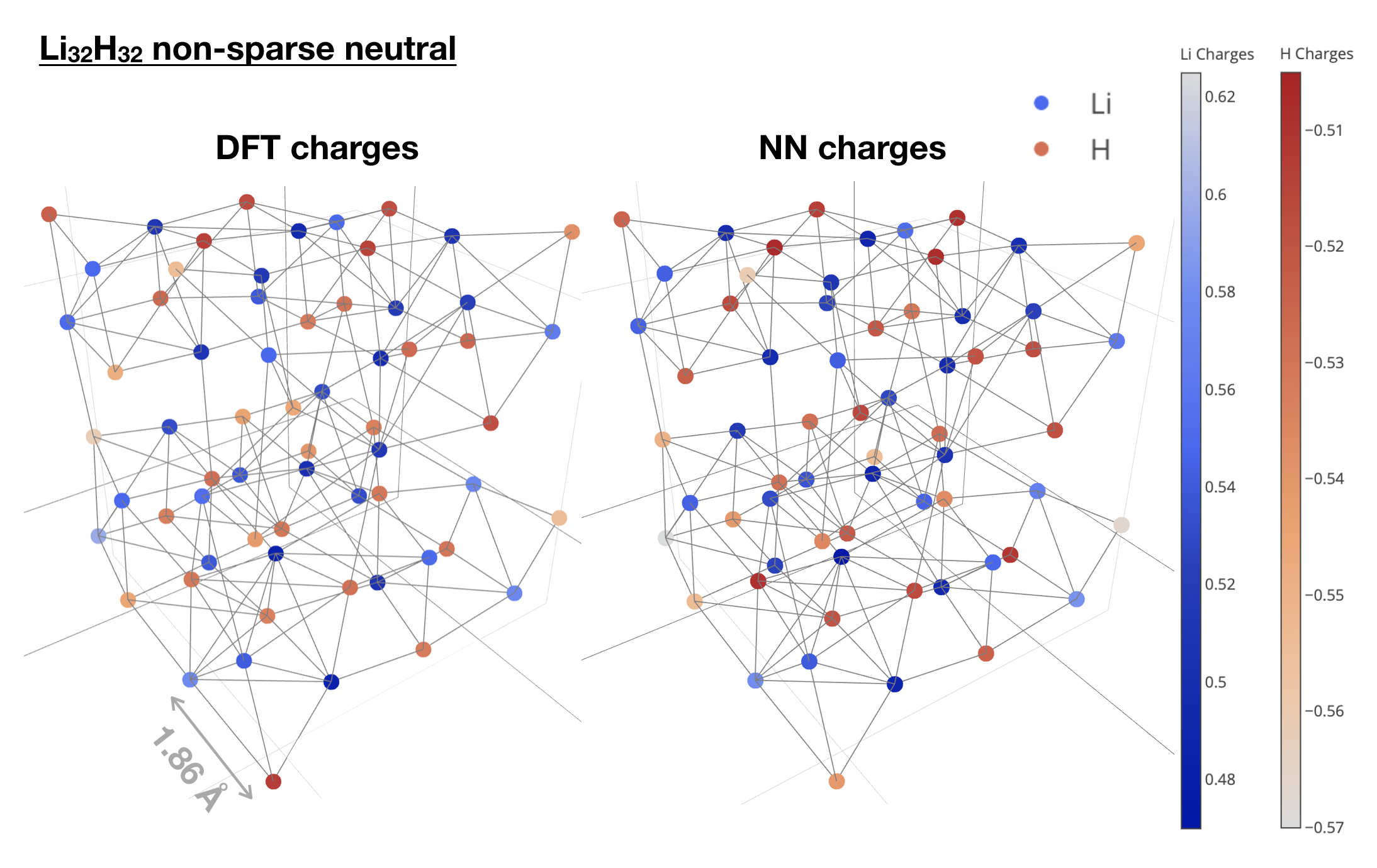}
     \caption{DFT and bpopNN partial charges for a neutral \ce{Li32H32}}
     \label{fig:subfig1}
  \end{subfigure}
  
  \begin{subfigure}[b]{0.4\linewidth}
    \includegraphics[width=\linewidth]{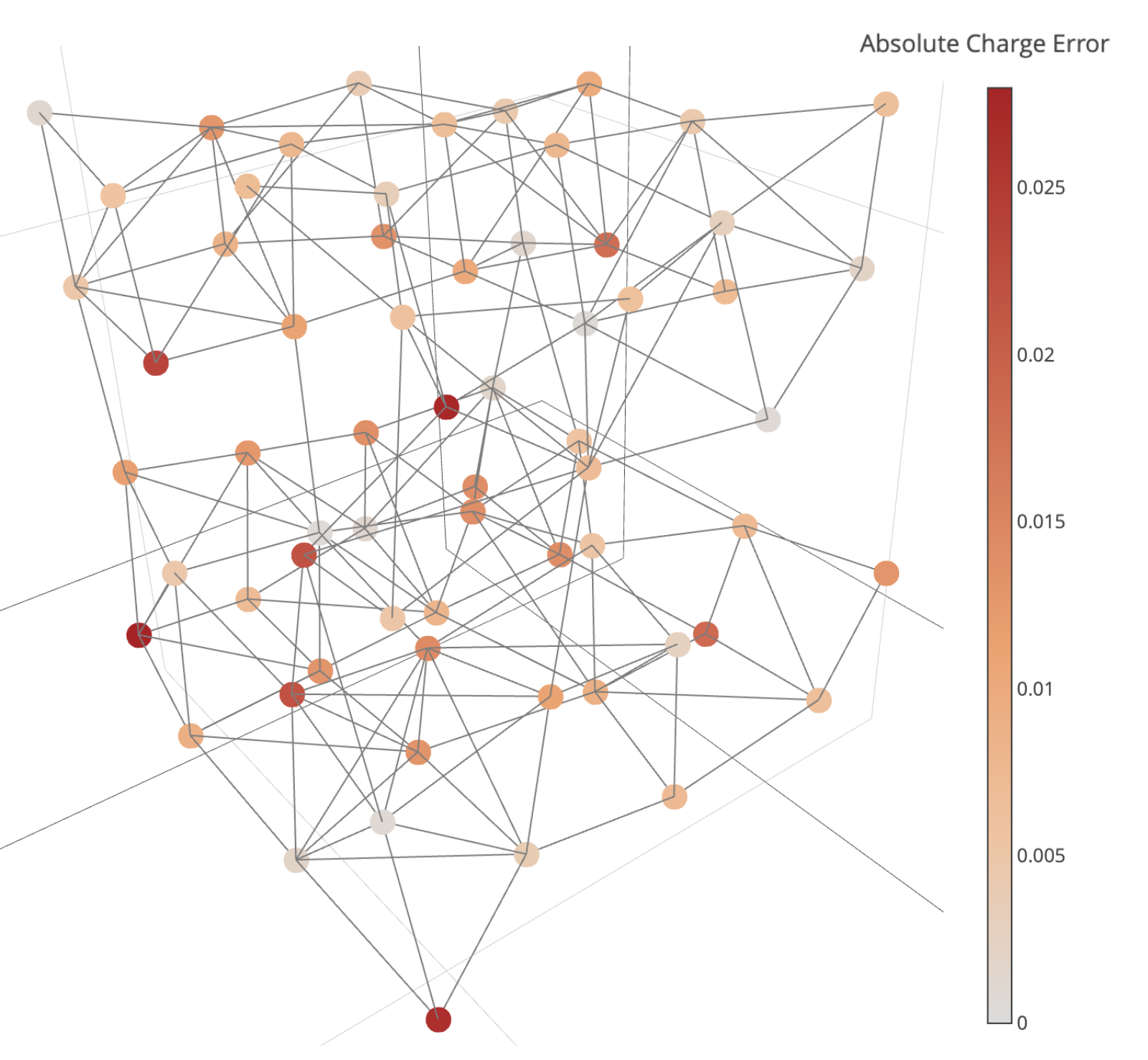}
     \caption{Charge error for a neutral \ce{Li32H32}}
     \label{fig:subfig2}
  \end{subfigure}
  \caption{Charge optimization result for a neutral \ce{Li32H32}}
  \label{fig:fig2}
\end{figure}

Figure ~\ref{fig:subfig1} compares the partial charges obtained from DFT and the \MLmodelName{} SCF-q procedure for a random test neutral \ce{Li32H32} structure; Figure ~\ref{fig:subfig2} depicts the absolute partial charge error for the same molecule. There is a good agreement between the two methods. The atoms on the corner of the molecule are showing slightly larger charge errors than the atoms on the edge or in the middle.

\begin{figure}[H]
    \centering
      \includegraphics[width=0.8\linewidth]{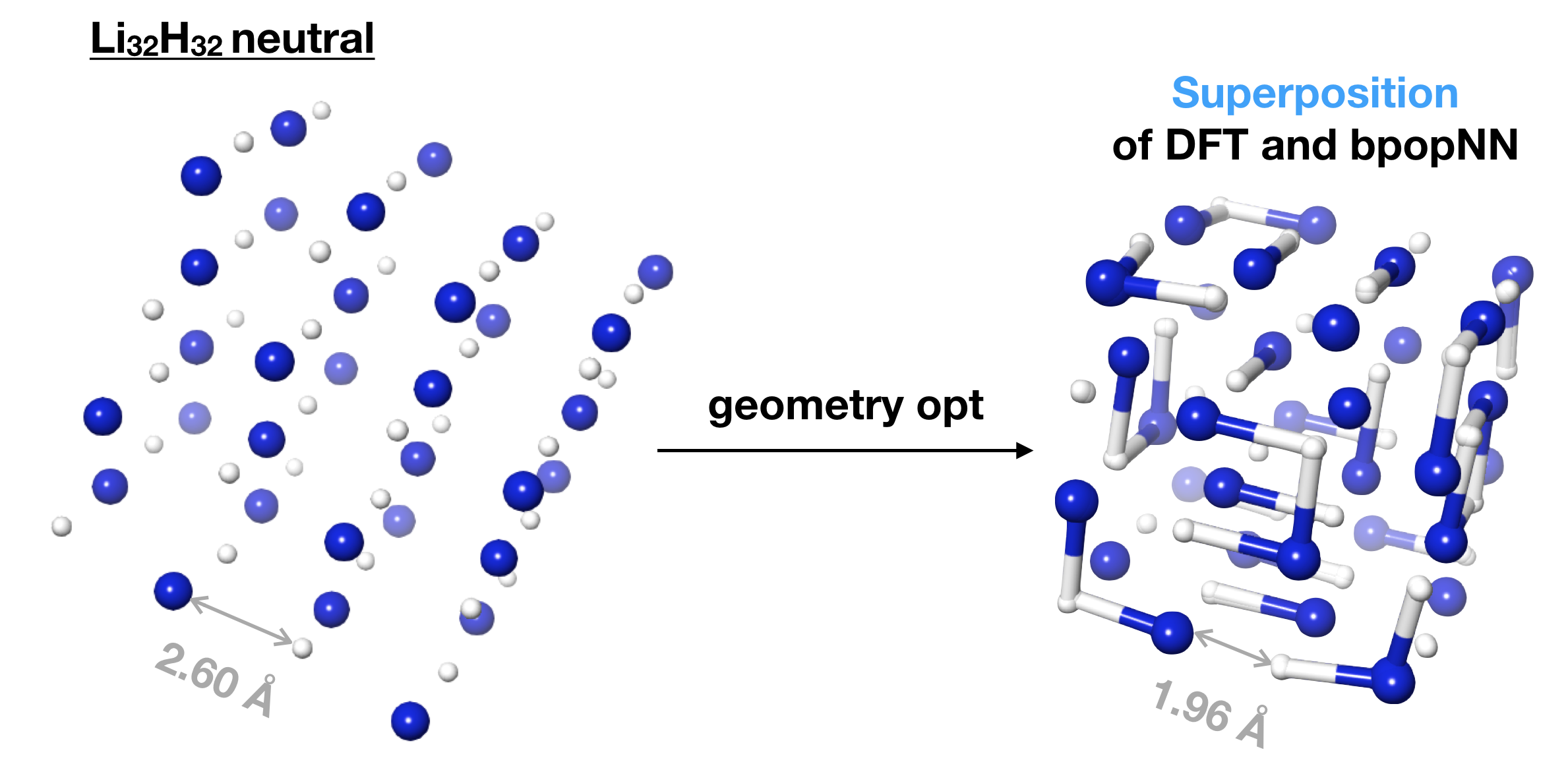}
     \caption{Geometry optimization on a cubic \ce{Li32H32} structure}
     \label{fig:fig4}
   \end{figure}

In addition to the SCF-q procedure, geometries and partial charges can be optimized simultaneously to a tight convergence ($10^{-7}$ Hartree/\AA{} for geometries and $10^{-7}$ Hartree/e for charges) using a conjugate gradient algorithm for both neutral and ionized molecules. We adopted the geometry-optimization procedure that DFT uses, i.e. converging SCF-q at each geometry cycle. In theory there is no need to follow this order, but this method turned out to be more stable and efficient in practice. The cubic \ce{Li32H32} structures were optimized using \MLmodelName{} separately for cation, neutral, anion, and compared with the DFT optimized geometries. Figure ~\ref{fig:fig4} shows the superposition of \MLmodelName{} and DFT optimized neutral \ce{Li32H32} structures. The equilibrium geometries were successfully predicted with an root mean square displacement (RMSD) of 0.0522 \AA{}, 0.0958 \AA{}, 0.0516 \AA{} for neutral, cation, and anion, respectively, compared to the true DFT geometries.  The superpositions and RMSD values were obtained with Maestro.\cite{maestro} The final absolute DFT and \MLmodelName{} energies are compared in Table ~\ref{table:table2}, along with the DFT and \MLmodelName{} predicted adiabatic ionization potential (IP) and electron affinity (EA) values. The predicted IP and EA are showing qualitative agreement with DFT (0.167 eV and 0.278 eV error for IP and EA respectively), although they are not highly accurate. These errors are fairly reasonable considering the size of the trial molecule (64 atoms) we are testing on, and the fact that B3LYP functional has $0.15$ to $0.18$ eV and $0.11$ to $0.16$ eV mean absolute deviations for IP and EA respectively for atoms and small main group molecules. \cite{ref56}

\subsubsection{Reaction Pathways}

\begin{figure}[H]
  \centering
  \includegraphics[width=\linewidth]{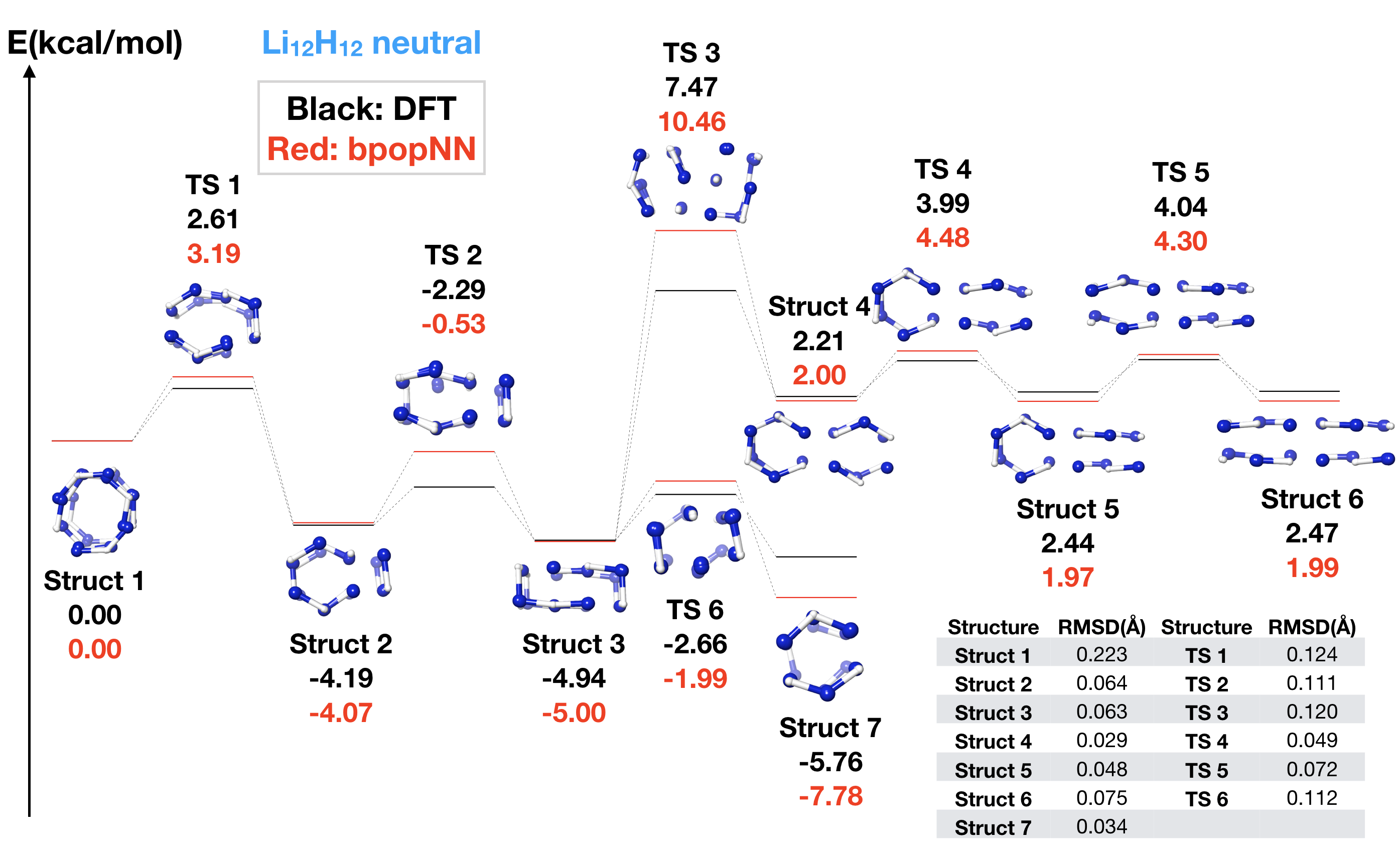}
  \caption{Energy profile for various structural transformations for 
neutral \ce{Li12H12}.  Superpositions and RMSD values were obtained with Maestro.\cite{maestro}}
  \label{fig:fig5}
\end{figure}

\begin{figure}[H]
  \centering
  \includegraphics[width=\linewidth]{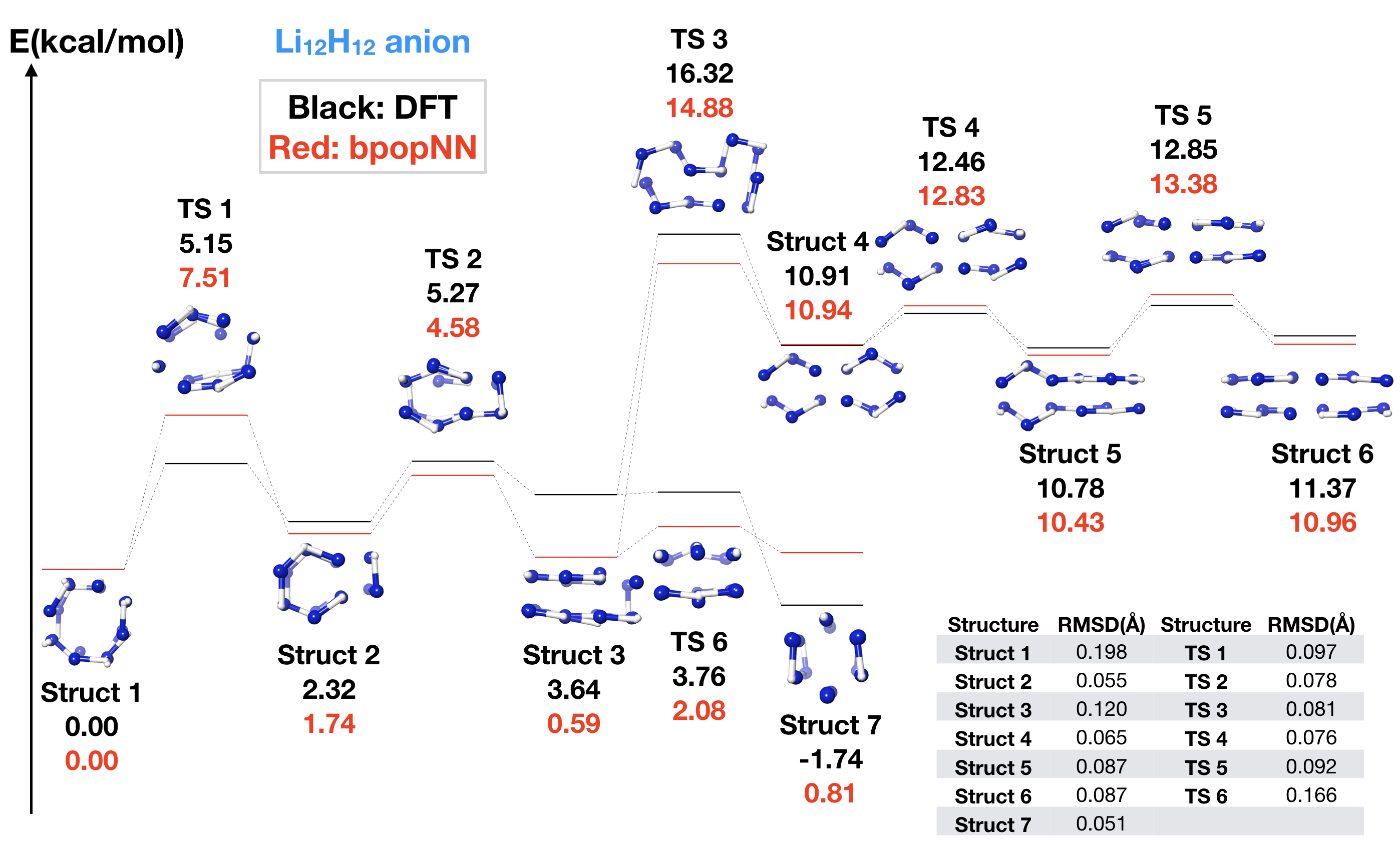}
  \caption{Energy profile for various structural transformations for 
anion \ce{Li12H12}.  Superpositions and RMSD values were obtained with Maestro.\cite{maestro}}
  \label{fig:fig6}
\end{figure}

As another practical application of our model, we sketched the energy landscape of various structural transformations for neutral and anionic \ce{Li12H12}. Diverse stationary-point structures were found for this system, ranging from cuboid structures to hexagonal structures to structures with fused 6-4-6 membered rings, etc. Although some motifs can be seen in a couple of optimized geometries in the training, these structures are overall distinct from the structures found in the training set. This provides an assessment of how well the model can perform in mapping out energetics for realistic reaction pathways.  The resulting energy profile for neutral \ce{Li12H12} structures is shown in Figure ~\ref{fig:fig5} and for anion \ce{Li12H12} in Figure ~\ref{fig:fig6}.  All structures shown were individually optimized, e.g.\ the DFT neutral and anionic structures are distinct.

The energies of \textbf{Struct 1} from DFT and \MLmodelName{} were separately used as baselines, so that only the energy difference between structures is shown.  The difference in energy between baseline structures for neutral relative to anion is 9.17 kcal/mol and 10.20 kcal/mol, respectively for DFT and \MLmodelName{}.
All the minimum structures and the transition states (TSs) connecting them were obtained independently for DFT and \MLmodelName{}. Newton's method was used in the \MLmodelName{} to converge the TSs (to geometry gradient thresh $10^{-7}$ Hartree/\AA{} and charge gradient thresh $10^{-7}$ Hartree/e) and the presence of one imaginary frequency was confirmed for all cases. The pathways from TSs to minimum structures in \MLmodelName{} were confirmed by a steepest descent algorithm with a small step size. All the DFT TSs were confirmed with one imaginary mode as well and pathways confirmed by Intrinsic Reaction Coordinate (IRC) calculations. In general, the \MLmodelName{} geometries show close agreement with DFT geometries (RMSD values are shown in the table inside each figure).  Energies for equilibrium structures are qualitatively good, although some deviations (2 to 3 kcal/mol) can be seen for \textbf{Struct 7} and anion \textbf{Struct 3}.  Reaction barriers are showing qualitative agreement as well, although there are cases with noticable quantitative error (neutral \textbf{Struct 3 $\leftrightarrow$ TS 3 $\leftrightarrow$ Struct 4}, neutral \textbf{Struct 3 $\leftrightarrow$ TS 6 $\leftrightarrow$ Struct 7}, anion \textbf{Struct 2 $\leftrightarrow$ TS 2 $\leftrightarrow$ Struct 3}). Some caution needs to be exercised if one were to predict the lowest energy structure from ML models, especially when the energy range is extremely small (See anion \textbf{Struct 1} vs \textbf{Struct 7}). 
At the very least, however, it is helpful to take advantage of the computational speed of the NN and map out the whole transformation picture prior to refinement with full electronic structure calculations.

\subsubsection{Long-range environmental effects}\label{long-range}

\begin{figure}[H]
  \centering
  \includegraphics[width=0.7\linewidth]{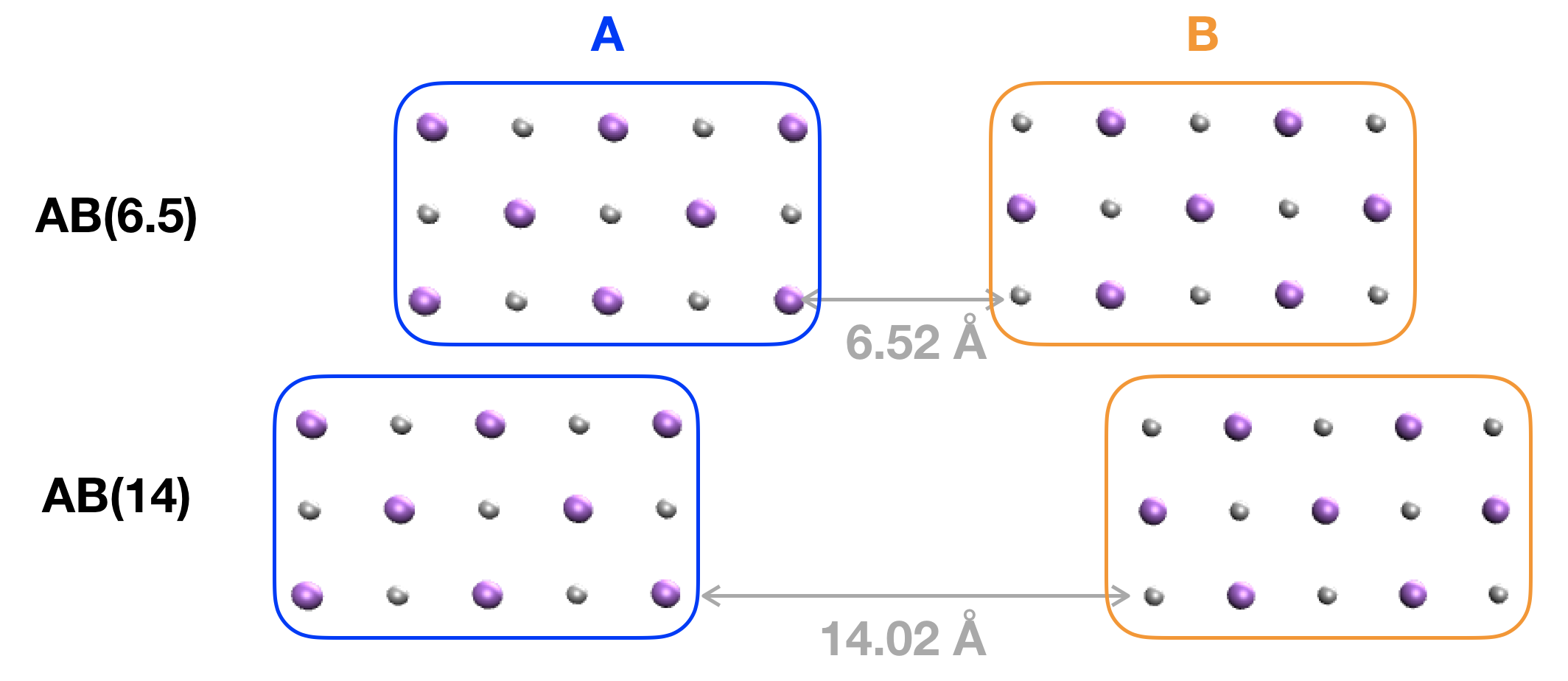}
  \caption{A charge separation example for neutral \ce{Li15H15}}
  \label{fig:fig7}
\end{figure}

Finally, we would like to demonstrate the importance of optimizing the electronic components self-consistently.  Namely, we consider a model system that exhibits significant long-range charge effects.  Consider a neutral \ce{Li15H15} cluster, which is the side-by-side interaction of two sheets, one Li$_8$H$_7$ (``A'') and one Li$_7$H$_8$ (``B'') as shown in Figure ~\ref{fig:fig7}.  The two sheets are separated by 6.52 \AA{} (``AB(6.5)'') or 14.02 \AA{} (``AB(14)'').  The first distance is beyond the cutoff used by other common local position-only descriptors and the second one is even beyond the long cutoff distance used in this work.  For B3LYP, both inter-sheet distances exhibit nearly full charge separation: the left sheet (A) bearing +1 charge and the right sheet (B) bearing -1 charge.  We note that a ML model based merely on local position-only descriptors will not succeed for this system.  In that case, for AB(14), monomer A doesn't ``see'' monomer B, so its predicted energy would mirror that for an isolated \emph{neutral} A monomer, and likewise for B.  The predicted atom charges and the intra-monomer energies would be incorrect, and the inter-monomer energy, which is largely electrostatic, would nearly be missed entirely.

As shown in Table ~\ref{table:table3}, this problem can be solved qualitatively by incorporating self-consistent electronic information: our model can predict the charge separation and qualitatively correct binding energies, although there were no direct scenarios of this kind in our training data.

At extremely large A-B separation, AB(1000), both DFT and \MLmodelName{} exhibit a partially ionized state.  For DFT, this stems from the well-known ``self interaction'' error (also called ``delocalization error''), and this kind of result is also expected of \MLmodelName{}.  For AB(6.5) and AB(14), \MLmodelName{} shows increased A-B charge separation relative to infinite separation.  The principal driving force for this, and the only one for AB(14), is the electrostatic attraction between A$^+$ and B$^-$.  In other words, the long-range environment is the sole reason for increased charge separation in \MLmodelName{} in this case, which explicitly establishes that the model is adaptable to the environment.

The sum of \MLmodelName{} energies for A$^+$ and B$^-$ is -121.4718 a.u., which actually is slightly above the sum of \MLmodelName{} energies for A and B, -121.4785 a.u.  The situation is opposite for DFT, which may partially explain the increased A-B charge separation for DFT for AB(6.5) and AB(14).

\section{Conclusions}

In this paper, we have highlighted the importance of incorporating adaptable electronic information into ML models of PESs.  For this, we presented a new general approach for building ML models whose target function is a DFT populations-to-energy map, afforded in practice by CDFT.  Such models are functions of atom positions and atom-based electron populations; optimization with respect to the latter variables allows for self-consistent adjustment to the electronic environment, akin to optimizing the density in DFT.  This leads to features analogous to regular orbital or density based SCF: solutions for different overall charge states; solutions with different character (e.g. CT and non CT); spin polarized
 versus not, etc.  In this sense, we view this as a reduced or simplified DFT approach. 
As a byproduct, this model can be used as a high-quality charge equilibration scheme, i.e.\ partial charges of this model can also be used by themselves to infer molecular properties, e.g.\ determining nucleophilic and electrophilic sites. 

As a proof of concept for this approach, we trained a \MLmodelName{} model for Li$_n$H$_n$ systems.  For this, we simplified the target-energy map to use only total electron populations on each atom, as opposed to the two spin populations.  Subsequent testing on a variety of Li$_n$H$_n$ examples confirmed that the model exhibits qualitative accuracy for the energy of different overall charge states, a flexible reduced electronic structure whose populations are generally close to those of the parent DFT functional and that can adapt to the (particularly medium to long-range) environment, and transferabilty to clusters of different sizes.  We conclude that \MLmodelName{} is a positive first step towards addressing some outstanding challenges for ML models, namely the handling of systems with electronic structure that is sensitive to the surrounding environment.

This first application was designed for qualitative testing and it has several aspects that can explain various inaccuracies and that are ready for refinement in future work.  Although partial charges are in principle adequate as a basis for the \MLmodelName{} approach, spin populations will need to be incorporated in practice to effectively model the intricacies of the CDFT map.  In this context, it should then be noted that the reduction of the density to atom-based populations is the furthest one can reasonably go; clearly less drastic reductions should also be explored, and perhaps a useful hierarchy for this can be established.

Improvements in the descriptor can be made.  Our descriptor is based on the ESP generated by the model's populations as point charges, and it uses simple Gaussian functions to approximate the ESP for each point.  We could better approximate these ESPs or incorporate them exactly.  It would be particularly interesting to combine the ideas of this paper with those of LODE:\cite{grisafi2019incorporating} instead of using point charges to generate the ESP, proxy densities, now integrating to the current populations on each atom, could be used to generate ESPs.

The fact that \MLmodelName{} models attempt to model true DFT partial charges is advantageous for the incorporation of long-range electrostatic interactions.  The simple approach used here, based on attenuated pairwise point-charge coulomb interactions, would clearly benefit from refinements to more reasonably approximate the local morphology of the interacting density pieces.

Experimenting with the choice of ML architecture would be worthwhile, and especially experimenting with the NN depth.  Of course, this would require a larger training dataset.  The dataset used here is on the smaller side, so enlarging this is a focal point for future work.  There is also the plausibility that our dataset is insufficiently balanced.  More specifically, it only contains alternating Li-H clusters and cannot be reasonably applied to clusters with distinguishable Li$_n$
and (H$_2$)$_n$ substructures, its geometries were obtained from trajectories using only the neutral charge state, and generally our sampling method of particular AIMD and optimization trajectories might not be varied enough.  In the future we will explore more efficient and balanced sampling methods, such as normal mode sampling. 

In the future, we are primarily interested in moving beyond the LiH-cluster scope to more general Li-organic systems.   This significant increase in complexity should entail many adjustments to the general approach presented above, and will serve as a firm test
of its efficacy.

\section{acknowledgement}
This work was supported as part of the Joint Center for Energy Storage Research (JCESR), an Energy Innovation Hub funded by the U.S. Department of Energy, Office of Science, Basic Energy Sciences, under Contract No. DE-AC02- 06CH11357. Calculations were performed at Lawrencium computational cluster, as well as the UC Berkeley Molecular Graphics and Computation Facility (MGCF).  MGCF is supported by grant NIH S10OD023532.

\begin{table}[H]
  \centering
  \caption{Energy and charge errors for \ce{Li29H29} sparse and \ce{Li32H32} non-sparse test sets}
  \label{table:table1}
  \begin{tabular}{c c c|c c c}
    \hline\hline
     & \makecell[c]{MAE charge \\ (e/atom)}  & \makecell[c]{MAE energy \\ (kcal/mol/atom)} & & \makecell[c]{MAE charge \\ (e/atom)} & \makecell[c]{MAE energy \\ (kcal/mol/atom)} \\
    \hline
     \makecell[c]{\ce{Li29H29} sparse \\ neutral} & 0.009 & 0.188 & \makecell{\ce{Li32H32} non-sparse \\ neutral} & 0.007 & 0.235 \\
    
     \makecell[c]{\ce{Li29H29} sparse \\ cation} & 0.011 & 0.136 & \makecell{\ce{Li32H32} non-sparse \\ cation} & 0.010 & 0.281 \\
    
     \makecell[c]{\ce{Li29H29} sparse \\ anion} & 0.013 & 0.121 & \makecell{\ce{Li32H32} non-sparse \\ anion} & 0.013 & 0.168 \\     
     
    \hline
  \end{tabular}
\end{table}

  \begin{table}[H]
  \centering
  \caption{Energy statistics for a cubic \ce{Li32H32} after geometry optimization}
  \label{table:table2}
  \begin{tabular}{c c c c}
    \hline\hline
     & \makecell[c]{DFT energy \\ (Hartree)}  & \makecell[c]{\MLmodelName{} energy \\ (Hartree)} & \makecell[c]{Energy Error \\ (kcal/mol/atom)}\\
    \hline
     Neutral & -260.956 & -260.967 & 0.106 \\
   
     Cation & -206.711 & -260.716 & 0.046 \\
    
     Anion & -260.962 & -260.983 & 0.206 \\     
     \hline
     \hline
     & DFT (eV) & \MLmodelName{} (eV) & \\
     \hline
     IP & 6.656 & 6.823 \\
     EA & 0.172 & 0.450 \\
       
    \hline
  \end{tabular}

\end{table}  
  
\begin{table}[H]
  \centering
  \caption{A charge separation example for neutral \ce{Li15H15}}
  \label{table:table3}
  \begin{tabular}{c c c c}
    \hline\hline
     & \makecell[c]{DFT energy \\ (Hartree)}  & \makecell[c]{\MLmodelName{} energy \\ (Hartree)} &  \makecell[c]{Absolute error \\ (kcal/mol)} \\
    \hline
     AB(6.5) & -121.524\textsuperscript{\emph{a}} & -121.534 & 6.232 \\
    
     AB(14) & -121.506\textsuperscript{\emph{a}} & -121.525 & 11.801 \\

     AB(1000) & -121.489\textsuperscript{\emph{a}} & -121.515 & 16.286 \\
    
     A & -64.213 & -64.224 & 7.213 \\     
     
     B & -57.241 & -57.254 & 8.403 \\

     A$^+$ & -64.110 & -64.112 & 1.210 \\     
     
     B$^-$ & -57.372 & -57.360 & 7.584 \\
     
     \hline\hline
     & DFT (kcal/mol) & \MLmodelName{} (kcal/mol) \\
     \hline 
     \makecell[c]{AB(6.5) BE\textsuperscript{\emph{b}}} & -44.143 & -34.760 \\
     
     \makecell[c]{AB(14) BE} & -33.168 & -29.345 \\
     
     \hline\hline
     & DFT (e) & \MLmodelName{} (e) \\
     \hline
     \makecell[c]{$q$(A)\textsuperscript{\emph{c}} in AB(6.5)} & 0.993 & 0.729 \\
     
     \makecell[c]{$q$(A) in AB(14)} & 0.979 & 0.681 \\ 

     \makecell[c]{$q$(A) in AB(1000)} & 0.683 & 0.609 \\
     
    \hline
  \end{tabular}
\end{table}
\textsuperscript{\emph{a}} For CT solutions. There exist DFT ``neutral-neutral'' solutions, with energies of -121.4542 and -121.4536 respectively for AB(6.5) and AB(14). This is very close to the sum of the energies of A and B. \\
\textsuperscript{\emph{b}} Binding Energy. \\
\textsuperscript{\emph{c}} sum of partial charges on A atoms.

\bibliography{main.bbl}

\appendix
\section{Population gradient of CDFT energy}
In the present context, the CDFT Lagrangian takes the following form:
\begin{equation}\label{cdft_lag}
L_{\text{CDFT}} = E[\gamma] + \sum_{\sigma} \sum_{i=1}^{N_{\text{atom}}-1} \lambda_{i,\sigma} (p_{i,\sigma}(\rho) - c_{i,\sigma}),
\end{equation}
where $\gamma$ is the density matrix, the $\lambda_{i,\sigma}$ are Lagrange multipliers, the $p_{i,\sigma}(\rho)$ are the functions defined in eqn.\ \eqref{basicBeckePop}, and
the $c_{i,\sigma}$ are the target values of the constraints.  
For each spin, the sum excludes one atom because the value of this last variable is implied 
by the usual condition that the density integrates to the total number of electrons of that spin.  An equivalent result is obtained if a different atom is the 
one excluded from the sum,
although of course the $\lambda_i$ and $c_i$ will adjust accordingly.  This Lagrangian also formally applies to cases in which fewer population constraints
are used.  In that case, we simply set the $c_i$ for any technically unconstrained population to its relaxed value, and the associated $\lambda_i$ is 0.

Optimizing $L_{\text{CDFT}}$ with respect to the orbitals (i.e.\ $\gamma$) and the Lagrange multipliers gives the CDFT energy,
$E_{\text{CDFT}}[\{c_{i,\sigma}\}]$.
At a stationary point of $L_{\text{CDFT}}$, we may apply the ``envelope'' theorem to obtain
\begin{equation}
\frac{\partial E_{\text{CDFT}} }{ \partial c_{i,\sigma} } = \frac{\partial L_{\text{CDFT}} }{ \partial c_{i,\sigma} } = -\lambda_{i,\sigma}.
\end{equation}

The target energy function $E_t[\bm{p}]$ is essentially the same as $E_{\text{CDFT}}[\{c_{i,\sigma}\}]$ except that it is a function of all $2N_{\text{atom}}$ population variables.
So we have to do a transformation between variables.
Again excluding the last atom, we can fix all 
$c_{i,\sigma}$ but one, $c_{j,\sigma}$, and then vary $c_{j,\sigma}$.  Upon (re-)optimizing $L_{\text{CDFT}}$ for this new set of constraints, 
$p_{j,\sigma}$ changes identically as $c_{j,\sigma}$ does, and
$p_{N_{\text{atom}},\sigma}$ changes by the opposite amount, while all other populations are unchanged.  Denoting the direction corresponding to this overall change 
in populations by $\mathbf{e}_{j,\sigma}$, we thus have

\begin{equation}
\nabla E_t[\bm{p}] \cdot \mathbf{e}_{i,\sigma} = \frac{\partial E_{\text{CDFT}} }{ \partial c_{i,\sigma} } = -\lambda_{i,\sigma}.
\end{equation}

The vectors $\mathbf{e}_{i,\sigma}$ are (most of) the columns of a block diagonal matrix $W$ with two $N_{\text{atom}}$ by $N_{\text{atom}}$ blocks, one for each spin type.
That is, $W = W_s \oplus W_s$ with
\begin{equation}
W_s = 
\begin{bmatrix}
1       & 0      & \dots  & 0       & 1 \\
0       & 1      & \dots  & 0       & 1 \\
\vdots  & \vdots & \dots  & \vdots  & \vdots \\
0       & 0      & \dots  & 1       & 1 \\
-1      & -1     & \dots  & -1      & 1 \\
\end{bmatrix}
.
\end{equation}

The last column in $W_s$ corresponds to the fully symmetric direction of changing all populations (of one spin type) by the same amount.  Because the total numbers of each
spin are constant, the component of $\nabla E_t$ along this direction is 0.  We thus have

\begin{equation}
\nabla E_t \cdot W = \bm{\nu},
\end{equation}
where the latter vector contains the $-\lambda_i$ along with two 0's for the symmetric directions: 
\begin{equation}
\nu_i = 
\begin{cases}
-\lambda_{i,\alpha} & \mbox{if } i < N_{\text{atom}} \\
0                   & \mbox{if } i = N_{\text{atom}} \\
-\lambda_{i-N_{\text{atom}},\beta}  & \mbox{if } N_{\text{atom}} < i < 2 N_{\text{atom}} \\
0                   & \mbox{if } i = 2 N_{\text{atom}}
\end{cases}
.
\end{equation}

By inverting $W$ we readily obtain $\nabla E_t[\bm{p}]$ (in the regular population basis).

\end{document}